\documentstyle[twocolumn,prl,aps]{revtex}
\begin{document}
\draft
%%%%%%%%%%%%%
\title{Statistical Distribution for Particles
obeying Fractional Statistics}
%%%%%%%%%%%%%
\author{Yong-Shi Wu}
\address{Department of Physics, University of Utah,
Salt Lake City, UT 84112 }
%%%%%%%%%%%%%
\date{\today}
%%%%%%%%%%%%%
\maketitle
\begin{abstract}
We formulate quantum statistical
mechanics of particles obeying fractional
statistics, including mutual statistics,
by adopting a state-counting definition.
For an ideal gas, the most probable
occupation-number distribution interpolates
between bosons and fermions, and respects
a generalized exclusion principle except for
bosons. Anyons in strong magnetic field
at low temperatures constitute such a physical
system. Applications to the thermodynamic
properties of quasiparticle excitations in the
Laughlin quantum Hall fluid are discussed.
\end{abstract}
%%%%%%%%%%%%%

\begin{narrowtext}

Statistics is the distinctive
property of a particle (or elementary
excitation) that plays a fundamental role
in determining macroscopic or thermodynamic
properties of a quantum many-body system.
For example, superfluidity or superconductivity
is essentially due to Bose-Einstein
condensation; and the stability of macroscopic
matter\cite{Lieb} depends crucially on the
Fermi-Dirac statistics of electrons and protons.
In recent years, it has been recognized that
particles with ``fractional statistics''
intermediate between bosons and fermions
can exist in two-dimensional\cite{FS} or in
one-dimensional\cite{Yang,Hald1} systems.
Most of the study has been done in the
context of many-body quantum mechanics.
Despite calculations of certain
thermodynamic properties in a few
examples\cite{Yang,Virial,Hald2,Ouvry}
with the help of exact solutions,
general formulation of quantum statistical
mechanics (QSM) with a distribution that
interpolates between bosons
and fermions is still lacking.

A single-particle
quantum state can accommadate an
arbitrary number of identical bosons, while
no two identical fermions can occupy
one and the same quantum state (Pauli's
exclusion principle).  In QSM\cite{Huang},
this difference gives rise to
different counting of many-body states,
or different statistical weight $W$.
For bosons or fermions, the number of quantum
states of $N$ identical particles occupying
a group of $G$ states is, respectively,
given by
\begin{equation}
W_{b}= {(G+N-1)! \over N!~ (G-1)!}~,~~~
{\rm{or}}~~ W_{f}= {G! \over N!~ (G-N)!}~.
\label{counting}
  \end{equation}
A simple generalization and interpolation
is
  \begin{equation}
 W = {[G+(N-1)(1-\alpha)]!
\over N!~ [G-\alpha N-(1-\alpha)]!}~,
\label{gnrl}
 \end{equation}
with $\alpha=0$ corresponding to
bosons and $\alpha=1$ fermions.
Such an expression can be the starting point
of QSM for {\it intermediate} statistics
with $0<\alpha<1$. Let us first clarify
its precise meaning in connection
with ``occupation of single-particle
states'', and justify it as a new
definition of quantum statistics,
{\it \`{a} la} Haldane \cite{Hald1}.

Following ref. \cite{Hald1},
we consider the situations
in which the number $G_i$ of linearly
independent single particle (or
elementary excitation) states
of species $i$, confined to a finite
region of matter, is {\it finite and extensive},
i. e. proportional to the size of
the matter region in which the particle exists.
Now let us add more particles {\it with the
boundary conditions and size of the
condensed-matter region fixed}. The
$N$-particle wave function, when the
coordinates of the $N-1$ particles
and their species are held fixed,
can be expanded in a basis of wave
functions of the $n$-th particle.
The crucial point is that in the presence
of other particles, the number $d_{i}$ of
available single-particle states in this basis
for the $n$-th particle of species $i$
generally is no longer a constant, as given
by $G_i$; rather it may depend on
the particle numbers $\{ N_{i} \}$
of all species. This happens, for
example, when localized particle states are
{\it non-orthogonal}; as a result,
the number of available
single-particle states changes
as particles are added at fixed
size and boundary conditions.
Haldane\cite{Hald1} defined the statistical
interactions $\alpha_{ij}$ through the relation
 \begin{equation}
\Delta d_i = - \sum_{j}
\alpha_{ij}~ \Delta N_j ~,
\label{Haldstat}
 \end{equation}
where $\{ \Delta N_{j} \}$ is a set of
allowed changes of the particle numbers.
In the same spirit, but more directly
for the purposes of QSM,
we prefer to define the statistics,
including possible mutual statistics,
by counting\cite{comm2} the number of
many-body states at fixed $\{ N_i \}$:
 \begin{equation}
W = {\prod}_i ~ { [G_i + N_{i}-1 -
\sum_j \alpha_{ij}(N_j-\delta_{ij})]!
\over (N_i)!~ [G_i - 1
- \sum_j \alpha_{ij}(N_j-\delta_{ij})]! }~.
\label{mystat}
 \end{equation}
The statistics parameters $\alpha_{ij}$
must be {\it rational}, in order that
a thermodynamic limit can be achieved
through a sequence of systems with
different sizes and particle numbers.
If we restrict to only one species,
(\ref{mystat}) reduces to (\ref{gnrl}).
Note there is no periodicity in
so-defined statistics $\alpha$.

Now let us consider the simple case that
every single-particle state of species $i$
has the same energy $\varepsilon_{i}$.
An ideal gas, by definition, is a system
whose total energy is a simple sum
 \begin{equation}
E=\sum_{i} N_{i} \varepsilon_{i},
\label{energy}
 \end{equation}
where each term is linear in the particle
number $N_{i}$.
Defering the discussions about when eqs.
(\ref{mystat}) and (\ref{energy}) are obeyed,
let us first apply them to study QSM of
such ideal gas.
Following the standard procedure\cite{Huang},
one may consider a grand canonical ensemble
at temperature $T$ and with chemical
potential $\mu_{i}$ for species $i$.
According to the fundamental principles of QSM,
the grand partition function is given by
(with $k$ the Boltzmann constant)
  \begin{equation}
Z= \sum_{\{N_{i}\}} W(\{N_{i}\})~
\exp \{\sum_{i} N_{i} (\mu_{i} -\varepsilon_{i})/kT \}~.
\label{partit}
  \end{equation}
As usual, we expect that for very large
$G_{i}$ and $N_{i}$, the summand has a
very sharp peak around the set of most-probable
(or mean ) particle numbers $\{N_{i}\}$.
Using the Stirling formula
$\log N! = N \log (N/e)$, and
introducing the average ``occupation number''
defined by $n_{i} \equiv N_{i}/ G_{i}$,
we express $\log W$ as (with $\beta_{ij}
\equiv \alpha_{ij} G_{j}/G_{i}$)
 \begin{eqnarray}
&&\sum_{i} G_{i} \{
- n_{i} \log n_{i} - (1 - \sum_j \beta_{ij} n_{j})
\log (1 - \sum_j \beta_{ij} n_{j})
\nonumber\\
&+& [1+\sum_j (\delta_{ij}-\beta_{ij}) n_{j}]
\log [1+\sum_j (\delta_{ij}-\beta_{ij}) n_{j}] \}.
\label{entropy}
 \end{eqnarray}
The most-probable distribution of $n_{i}$
is determined by
 \begin{equation}
{\partial \over \partial n_{i}}\,
\bigl[ \log W + \sum_{i}
G_{i} n_{i}\,(\mu_{i} - \varepsilon_{i})/kT \bigr] =0~,
\label{equil}
 \end{equation}
It follows that
 \begin{eqnarray}
n_i~ e^{(\varepsilon_{i}-\mu_{i})/kT} & = &
[1+ \sum_{k} (\delta_{ik}-\beta_{ik}) n_k]
\nonumber\\
&\times & \prod_{j}~ \Bigl[ {1-\sum_{k}\beta_{jk} n_k \over
1+ \sum_k (\delta_{jk}-\beta_{jk}) n_k }\Bigr]^{\alpha_{ji}}~.
\label{eq1}
 \end{eqnarray}
Setting $w_i=n_i^{-1}-\sum_k \beta_{ik} n_k/n_i$,
we have
 \begin{equation}
(1+w_i)  \prod_{j} \Bigl({w_j
\over 1+w_j}\Bigr)^{\alpha_{ji}}
= e^{(\varepsilon_i-\mu_{i})/kT}.
\label{eqw}
 \end{equation}

Therefore the most-probable
average occupation numbers
$n_i (i=1,2,\cdots)$ can be obtained by
solving
 \begin{equation}
\sum_{j} (\delta_{ij}w_j +\beta_{ij}) n_j = 1~,
\label{eqn}
 \end{equation}
with $w_i$ determined by the
functional equations (\ref{eqw}).
The thermodynamic potential
$\Omega=-kT \log Z$ is given by
 \begin{equation}
\Omega \equiv - PV = -kT \sum_i G_i
\log {1+ n_i - \sum_j \beta_{ij} n_j
\over 1- \sum_j \beta_{ij} n_j}~;
\label{omega1}
 \end{equation}
and the entropy, $S=(E-\sum_{i}\mu_{i}N_{i}
-\Omega)/T$, is
 \begin{equation}
{S\over k}= \sum_i G_i
\Bigl\{ n_i {\varepsilon_i - \mu_{i} \over kT} +
\log {1+ n_i - \sum_j \beta_{ij} n_j
\over 1- \sum_j \beta_{ij} n_j } \Bigr\}.
\label{entropy2}
 \end{equation}
Other thermodynamic functions follow
straightforward\-ly. As usual, one can
easily verify that the fluctuations,
$({\overline{{N_{i}}^{2}}}-{\bar{N_{i}}}^{2})/
{\bar{N_{i}}}^{2}$, of the occupation numbers
are negligible, which justifies the validity of
above approach.

To see more explicitly the consequences of
the above results, consider the
simplest case of only one species
with different energy levels,
for which we set $\alpha_{ij}=
\alpha \delta_{ij}$ and $\mu_{i}=\mu$.
Then the average occupation number $n_i$
satisfies
 \begin{equation}
(1-\alpha n_{i})^{\alpha} [1+(1-\alpha) n_{i}]^{1-\alpha}
= n_{i} e^{(\varepsilon_{i}-\mu)/kT}~;
\label{eqn2}
 \end{equation}
and we have the statistical distribution
 \begin{equation}
n_{i} = { 1 \over w(e^{(\varepsilon_{i}-\mu)/kT})
+ \alpha}~,
\label{distr2}
 \end{equation}
where the function $w(\zeta)$ satisfies the functional
equation
 \begin{equation}
w(\zeta)^{\alpha} [1+ w(\zeta)]^{1-\alpha} = \zeta
\equiv e^{(\varepsilon-\mu)/kT}~.
\label{eqw2}
 \end{equation}

Note that $w(\zeta)= \zeta-1$ for $\alpha=0$;
and $w(\zeta)= \zeta$ for $\alpha=1$. Thus,
eq. (\ref{distr2}) recovers
the familiar Bose and Fermi distributions,
respectively, with $\alpha=0$ and $\alpha=1$.
For semions with $\alpha=1/2$,
eq. (\ref{eqn2}) becomes a quadratic equation,
which can be easily solved to give
 \begin{equation}
n_{i} = {1 \over
\sqrt{ 1/4 + \exp [2(\varepsilon_{i}-\mu)/kT]}}~.
\label{semion}
 \end{equation}
For intermediate statistics
$0< \alpha <1$, it is not hard to
select the solution $w(\zeta)$
of eq. (\ref{eqw2}) that interpolates
between bosonic and fermionic distributions.
In particular, when $\zeta$ is very large,
we have $w(\zeta)\approx \zeta$ and, neglecting
$\alpha$ compared to $w(\zeta)$, we recover the
Boltzmann distribution
 \begin{equation}
n_{i} = e^{-(\varepsilon_{i}-\mu)/kT},
\label{Boltz}
 \end{equation}
at sufficiently low densities for any statistics.

Furthermore, we note that $\zeta$ is
always non-negative, so is $w$;
it follows from eq. (\ref{distr2}) that
 \begin{equation}
n_{i} \leq 1/\alpha~.
\label{pauli}
 \end{equation}
This expresses the generalized exclusion
principle for fractional statistics.
In particular, at absolute zero, $\zeta=0$
if $\varepsilon_i< \mu$, and $\zeta=+\infty$
if $\varepsilon_i>\mu$. From eq. (\ref{eqw2}),
we have $w=0$ and $\infty$ respectively.
Thus, we see that at $T=0$, for statistics
$\alpha\neq 0$, the average occupation numbers
for single-particle states with continuous
energy spectrum obey a step distribution
like fermions:
 \begin{eqnarray}
n_i = &0, \;\;\;\; &\rm{if}~~  \varepsilon_i>E_F;
\nonumber\\
      &1/\alpha, \;\;\;&\rm{if}~~ \varepsilon_i<E_F.
\label{step}
 \end{eqnarray}
The Fermi surface $\varepsilon=E_F$
is determined by the requirement
$\sum_{\varepsilon_i<E_F} G_i=\alpha N$.
Below the Fermi surface, the average
occupation number is $1/\alpha$ for
each single-particle state,
a consequence of generalized
exclusion principle.

One may be tempted to consider,
in parallel to usual Bose and Fermi
ideal gas, the case\cite{comm3} with
 \begin{equation}
\varepsilon_i= {\hbar^2 k_{i}^2 \over 2m},~~
G_i= {V k_{i}\Delta k_{i}\over 2\pi}~,
\label{hypoth}
 \end{equation}
say in two dimensions with $V$ the area.
Then treating momentum as continuous,
one has the Fermi momentum
 \begin{equation}
k_F^2= (4\pi \alpha) (N/V).
\label{fermi}
 \end{equation}
Moreover, at finite temperatures,
by using (\ref{distr2}) and (\ref{eqw2}),
the sum $\sum_{i} G_{i} n_{i}=N$ can
be done to give
 \begin{equation}
{\mu \over kT} = \alpha {2\pi \hbar^{2}\over mkT} {N\over V}
+ \log \Bigl[ 1- \exp \Bigl(
- {2\pi\hbar^{2}\over mkT} {N\over V} \Bigr) \Bigr]~.
 \end{equation}
Using the identity derived by integration
by parts,
 \begin{equation}
\int_{0}^{\infty} d\varepsilon_{i}
\log { 1- \alpha n_{i} \over 1 + (1-\alpha) n_{i}}
= \int_{0}^{\infty} d\varepsilon_{i} \varepsilon_{i} n_{i}~,
\label{identity}
 \end{equation}
we have the statistics-independent
relation $PV=E$.
In the Boltzmann limit ($\exp(\mu/kT)
<< 1$), $w(\zeta)=\zeta+\alpha-1$,
 \begin{equation}
PV = NkT [ 1+
(2\alpha -1) N \lambda^{2}/4 V ]~,
 \end{equation}
where $\lambda=\sqrt{2\pi\hbar^2/mkT}$.
So the ``statistical interactions'' are
attractive or repulsive depending on
whether $\alpha<1/2$ or $\alpha >1/2$.

Whether a given system satisfies
the seemingly harmless conditions
(\ref{energy}) or (\ref{hypoth})
together with (\ref{mystat})
is a nontrivial question. The
state-counting definition (\ref{mystat})
for fractional statistics does not apply
\cite{Hald1} to free anyons, i.e. Newtonian particles
carrying flux-tubes\cite{Wil,Wu}.
On the other hand, anyons in magnetic
field satisfy (\ref{mystat}), but not
the condition (\ref{energy}) unless (at
very low temperatures when) all anyons
are in the lowest Landau level (LLL)
\cite{magnet}.
The Jastrow-type prefactor
$\Pi_{a<b}(z_a-z_b)^{\theta/\pi}$, with
$0\leq\theta <2\pi$, in the anyon wave function
has the effect of increasing the flux through the
system by $(\theta/\pi) (N-1)$. Thus, with
fixed size and number of flux,
the dimension of the effective boson Fock space
\cite{Hald3,Zhang} is given by
$d = N_{\phi} - (\theta/\pi) (N-1)$,
where $N_{\phi}= qBV/hc\equiv V/V_{0}$,
with $q$ the anyon charge. Eq. (\ref{gnrl})
applies, with the single-anyon
degeneracy $G=N_{\phi}$ and the statistics
$\alpha=\theta/\pi$. Applying eqs. (\ref{distr2})
and (\ref{eqw2}) with only one energy
$\varepsilon= \hbar\omega_c/2$, we have
 \begin{equation}
n\equiv {N\over G}
\equiv {\rho \over \rho_{0}}=
{1 \over w( e^{(\varepsilon-\mu)/kT})+\alpha}~,
\label{eqn3}
 \end{equation}
where $w(\zeta)$ is the positive solution
of eq. (\ref{eqw2}). Here
$\rho\equiv N/V$ is the areal density, and
$\rho_{0}\equiv 1/V_{0}$.
Eq. (\ref{eqn3}), together with (\ref{eqw2}),
determines the chemical potential $\mu$ in terms
of $\rho/\rho_{0}$ and T. Thermodynamic
quantities can also be expressed
as functions of the ratio $\rho/\rho_{0}$.
In particular, the thermodynamic potential is
 \begin{equation}
\Omega= -kT{V \over V_0}
\log {1+w \over w}
= -kT {V \over V_0} \log
{1+ (1-\alpha) n \over 1-\alpha n}.
\label{poten2}
 \end{equation}
The equation of state is
 \begin{equation}
{PV \over NkT} =
\Bigl({\rho \over \rho_0} \Bigr)^{-1}
\log {1+(1-\alpha)(\rho/\rho_0)
\over 1-\alpha (\rho/\rho_0)}~.
\label{state}
 \end{equation}
The pressure $P$ is linear in $T$
for fixed $\rho$. It diverges
at the critical density $\rho_{c}
=(1/\alpha)\rho_{0}$, which corresponds to
the complete filling of the LLL.
The emergence of an incompressible
state at filling fraction $1/\alpha$
is a consequence of the generalized
exclusion principle (\ref{step}).
\cite{commf}
The magnetization per unit area is
 \begin{equation}
{\cal M}
= - \mu_0 \rho +
{2\mu_0 \over \lambda^2}
\log {1+ (1-\alpha)(\rho/\rho_0)
\over 1-\alpha (\rho/\rho_0)},
\label{tization}
 \end{equation}
where $\mu_0=q\hbar/2mc$ is the Bohr magneton.
Note the first (de Haas-van Alphen) term
is statistics-independent. At low temperatures,
$kT << \hbar \omega_c$, the second term can be
neglected except for $\rho$ very close
to $(1/\alpha)\rho_{0}$, where it gives rise to
a non-vanishing, $\alpha$-dependent susceptibility
 \begin{equation}
\chi
= kT {q \over 2\pi \hbar c}
\bigl(- {1 \over B}\bigr)
{\rho/\rho_{0} \over (1-\alpha \rho/\rho_{0})
[1+(1-\alpha)\rho/\rho_{0}]}~.
\label{suscept}
\end{equation}
The entropy per particle is also $\alpha$-dependent:
 \begin{eqnarray}
{S \over N} = && k (1-\alpha +
{\rho_{0} \over \rho })
\log [1+(1-\alpha) {\rho \over \rho_{0} }]
\nonumber\\
&& - k\log {\rho \over \rho_{0} }
- k({\rho_{0} \over \rho }-\alpha)
\log (1-\alpha {\rho \over \rho_{0}})~.
\label{entropyr}
 \end{eqnarray}
Eqs. (\ref{poten2}), (\ref{state})
and (\ref{tization}) have been
derived in Ref.\cite{Ouvry} from the known
exact many-anyon solutions in the LLL.

Vortexlike quasiparticle excitations
in Laughlin's incompressible $1/m$-fluid
(m being odd)\cite{Laughlin} are known to be
fractionally charged anyons, and their
wave functions are such as if they are in
the LLL (with electrons acting as
quantized sources of ``flux'')
\cite{Hald3}-\cite{Johnson}.
The existence of two species of excitations,
quasi-holes (labeled by $-$) and quasi-electrons
(labeled by $+$), dictates nontrivial mutual statistics.
{}From the ``total-flux'' constraint
 \begin{equation}
N_{\phi}\equiv eBV/hc = mN_{e} + N_{-} - N_{+}~,
\label{flux}
 \end{equation}
where $N_{e}$ the number of electrons,
it follows\cite{Hald1} that
 \begin{equation}
\alpha_{--}=-\alpha_{+-}=\alpha_{-+}=1/m,~~~
\alpha_{++}=2-1/m~,
\label{value}
 \end{equation}
(appropriate for {\it hard-core}
quasielectrons\cite{Zhang,Johnson}).
The single excitation degeneracy
in the thermodynamic limit
is $G_{+}=G_{-}= (1/m)N_{\phi}$.
Ignoring the interaction energies
and assuming the system is pure,
we apply the formulas
(\ref{eqw}-\ref{entropy2}) to this case
with the values (\ref{value}) for $\alpha_{ij}$.
The densities $\rho_{\pm}$
of the excitations are given by
 \begin{equation}
n_{\pm}= {\rho_{\pm} \over \rho_{0}}=
{ w_{\mp}+\alpha_{\mp\mp}-\alpha_{\pm\mp}
\over (w_{+}+\alpha_{++}) (w_{-}+\alpha_{--})
-\alpha_{+-}\alpha_{-+} }~.
\label{eqn4}
 \end{equation}
where $\rho_{0}\equiv G_{\pm}/V$;
$w_{\pm}$ satisfy the functional equations
 \begin{equation}
{w_{\pm}}^{\alpha_{\pm\pm}} (1+w_{\pm})^{1-\alpha_{\pm\pm}}
\Bigl( {w_{\mp} \over 1+w_{\mp}} \Bigr)^{\alpha_{\mp\pm}}
=  e^{(\varepsilon_{\pm}-\mu_{\pm})/kT }~.
\label{eqw4}
 \end{equation}
Here $\varepsilon_{\pm}$ is the creation energy of a single
excitation. At $T=0$ or very close to it, thermal
activation is negligible; there is only one species of
excitations, behaving like anyons in the LLL.
Say, if $N_{\phi}<mN_{e}$, then there are only
quasielectrons: $n_{-}=0$,
$n_{+}=(mN_{e}-N_{\phi})/G_{+}=1/(w_{+}+\alpha_{++})$.
One can apply the above eqs. (\ref{eqn3})-(\ref{entropyr}).
At higher temperatures, thermal activation of
quasiparticle pairs, satisfying $\mu_{+}+\mu_{-}=0$,
becomes important and the effects of mutual
statistics become manifest with increasing density
of activated pairs. The thermodynamic
properties at different sides of electron filling
$\nu\equiv N_{e}/N_{\phi}=1/m$ are not symmetric
due to asymmetry in quasielectrons
($\alpha_{++}=2-1/m$) and quasiholes ($\alpha_{--}=1/m$).
The general equation of state is
 \begin{equation}
{P\over kT}
= \rho_{0} \sum_{i=+,-} \log {1+ (\rho_{i} / \rho_{0})
- \sum_{j} \alpha_{ij} (\rho_{j} / \rho_{0})
\over 1 - \sum_{j} \alpha_{ij} (\rho_{j} / \rho_{0}) }~.
 \end{equation}
When the excitation densities satisfy
 \begin{equation}
\sum_{j=+,-} \alpha_{ij} \rho_{j}= \rho_{0}~,
\label{plat}
 \end{equation}
the pressure diverges and a new incompressible
state is formed, as a result of the generalized
exclusion principle obeyed by the
excitations upon completely filling the LLL.
At $T=0$, e.g.,
for $i=+$ (quasiparticles) the limit
(\ref{plat}) is reached when electron filling
is $\nu=2/(2m-1)$, giving rise to the
well-known hierarchial state\cite{Hald3,Ma,Zhang}.
At finite $T$, it may happen at somewhat
different filling, because of the additional
quasihole contribution in eq. (\ref{plat}).
Moreover, the magnetization per unit area is
 \begin{equation}
{\cal M}
= \sum_{i} \Bigl( -\mu_{i} \rho_{i}
+ {e kT \over m hc}
\log { \rho_{0} + \rho_{i}
-\sum_{j} \alpha_{ij} \rho_{j}
\over \rho_0 -\sum_{j} \alpha_{ij} \rho_{j}}\Bigr).
\label{tization2}
 \end{equation}
Here $\mu_{\pm}=\partial \varepsilon_{\pm}/\partial B$;
the linearity of $\varepsilon_{\pm}$ in $B$ is
observed in recent experiments\cite{Tsui}.
Hopefully, when $kT$ is of order of $\varepsilon_{\pm}$
or higher, the $\alpha_{ij}$-dependent second
term may give an appreciable contribution.
The mutual-statistics-dependent
total entropy, $S=\sum_{i}N_{i}s_{i}$,
is
 \begin{eqnarray}
&&{s_{i}\over k} = - \log {\rho_{i}\over \rho_{0}}
- ({\rho_{0}\over \rho_{i}}
- \sum_j \alpha_{ij} {\rho_{j}\over \rho_{i}})
\log (1 -\sum_j \alpha_{ij} {\rho_{j}\over \rho_{0}})
\nonumber\\
&+& [1 + {\rho_{0}\over \rho_{i}} -
\sum_j \alpha_{ij} {\rho_{j}\over \rho_{i}}]
\log [1+\sum_j (\delta_{ij}-\alpha_{ij}) {\rho_{j}\over \rho_{0}}].
\label{entropy4}
 \end{eqnarray}

To conclude, we have formulated the QSM of
particles of fractional statistics (and mutual
statistics), adopting a state-counting
definition. When their interaction
energies can be neglected,
they obey a statistical distribution that
interpolates between bosons and fermions,
and respect a generalized exclusion principle
which makes them behave more or less like
fermions.  Anyonic quasiparticle excitations
in the Laughlin quantum Hall liquids are such
particles with nontrivial mutual statistics.
It would be interesting to experimentally
test their thermodynamic properties. Also it
would be interesting to see if and how statistics
with $\alpha \geq 2$, which makes perfect sense with
the state-counting defintion (\ref{gnrl}),
realizes in Nature.

I thank Fu-Chun Zhang for helpful
discussions. This work was supported
in part by NSF grant PHY-9309458.

\end{narrowtext}


\begin{references}
\bibitem{comm2}
At fixed particles numbers, the particles
of species $i$ can be regarded as
either bosons with a Fock space dimension
$d_B= d_i$, or fermions with a
Fock space dimension $d_F=d_i +N_i -1$.
Moreover, $d_{i}= G_i - \sum_j
\alpha_{ij}(N_j-\delta_{ji})$ solves
(\ref{Haldstat}) under the condition that
$d_i=G_i$ when $N_j=\delta_{ji}$.
\bibitem{comm3}
We assume the system has a natural cutoff at large
momenta so that the total number of single-particle
states is finite. But we will neglect the effect of
a finite cutoff in various relevant integrals.
\bibitem{commf}
We note that the Laughlin wave function for the
$1/m$-state looks the same as if the LLL is completely
filled by ``super''-electrons with statistics $\alpha=m$.
\end{references}
\end{document}